\documentclass[11pt]{article}
\pdfoutput=1

\usepackage[margin=1in]{geometry}
\usepackage{amsmath,amssymb}
\usepackage{graphicx}
\usepackage{booktabs}
\usepackage[colorlinks=true,linkcolor=blue,citecolor=blue,urlcolor=blue]{hyperref}
\usepackage{url}
\usepackage{listings}
\usepackage{xcolor}

\hypersetup{pdftitle={peapods: A Rust-Accelerated Monte Carlo Package for Ising and XY Spin Systems}, pdfauthor={Yan Ru Pei}}
\newcommand{\pkg}{\texttt{peapods}}
\newcommand{\code}[1]{\texttt{#1}}

\title{\texttt{peapods}: A Rust-Accelerated Monte Carlo Package\\for Ising and XY Spin Systems}
\author{Yan Ru Pei\\\texttt{yanrpei@gmail.com}}
\date{}

\begin{document}
\maketitle

\begin{abstract}
We present \pkg{}~\cite{Peapods}, an open-source Python package for Monte Carlo simulation of classical spin models with arbitrary couplings, whose sampling loop runs in Rust. It samples Ising models on periodic Bravais lattices given by user-specified neighbor offsets, optionally with open or antiperiodic boundaries, and XY models with signed couplings and site dilution on hypercubic lattices. Besides Metropolis and heat-bath sweeps, Swendsen--Wang and Wolff cluster updates and parallel tempering, it provides a family of replica cluster moves for spin glasses: the Houdayer, J\"org and Chayes--Machta--Redner moves, multi-pair variants of the first two, replica-permutation moves and a gauged form of them, and Swendsen--Wang replica Monte Carlo across temperatures. The same-temperature moves share one construction, a symmetry of the replicas' joint energy applied on clusters, and one-step stationarity tests from exact Boltzmann samples on small lattices check each family of moves that satisfies detailed balance. The package measures overlap distributions, spin-glass correlation lengths, autocorrelation times, equilibration and tempering-flow diagnostics, and jackknife errors over disorder samples. It reproduces the exact critical temperatures of the square and triangular Ising lattices, and its Binder curves for simple-cubic, BCC and FCC lattices agree at the literature critical temperatures. For the three-dimensional $\pm J$ spin glass, the crossings of $\xi_{\mathrm{SG}}/L$ for $L \le 12$ extrapolate, with the Janus correction exponent held fixed, to $T_c = 1.085(42)$, consistent with the Janus value $1.1019(29)$. For the two-dimensional XY model, the helicity modulus, correlation length and structure factor agree with high-precision reference data within $1.9$ standard errors. This article describes version 0.3.0.
\end{abstract}

\section{Introduction}

The Ising model is a cornerstone of statistical mechanics and the study of phase transitions. With disordered couplings it becomes the Edwards--Anderson spin glass~\cite{Edwards1975}, whose low-temperature phase in finite dimensions is still debated. Monte Carlo simulation is the main numerical tool for both, and both stress it. Near a critical point, single-spin-flip dynamics decorrelates slowly, which cluster algorithms built on the Fortuin--Kasteleyn representation~\cite{FortunKasteleyn1972,Swendsen1987,Wolff1989} largely cure for ferromagnets. In spin glasses these clusters percolate above the spin-glass transition, so flipping them does little, and simulations rely instead on parallel tempering~\cite{Hukushima1996}, on cluster moves that act on several replicas at once~\cite{SwendsenWang1986,Houdayer2001,Jorg2006,RMC1998}, and on other non-local or non-equilibrium schemes~\cite{PeiDiVentra2022}. Which combination pays depends on the dimension, the lattice size and the temperature, so it helps to have all of them in one package with the diagnostics needed to compare them.

Python dominates scientific computing, but its interpreter overhead rules it out for the inner loops of Monte Carlo simulation. NumPy vectorization materializes large temporary arrays, Numba supports few complex data structures, and C extensions give up memory safety. \pkg{} instead writes the whole sampling loop in Rust, which compiles to code competitive with C and enforces memory safety at compile time, and exposes it to Python through PyO3~\cite{PyO3} and Maturin~\cite{Maturin}. Python handles model construction, analysis and plotting.

This article describes version 0.3.0. The package provides:
\begin{itemize}
  \item Ising models on any periodic Bravais lattice given by integer neighbor offsets, with open or antiperiodic boundaries along any axis, and XY models with signed couplings and site dilution on hypercubic lattices (Section~\ref{sec:models});
  \item Metropolis and heat-bath sweeps, Swendsen--Wang and Wolff cluster updates, an exact von Mises heat bath and overrelaxation for XY spins, and parallel tempering with a non-reversible full-ladder schedule (Sections~\ref{sec:single} and~\ref{sec:pt});
  \item replica cluster moves for spin glasses, the same-temperature ones built from one symmetry construction, each checked by a stationarity test against exact Boltzmann samples (Section~\ref{sec:overlap});
  \item spin-glass observables, autocorrelation times, equilibration and tempering diagnostics, and error estimates over disorder samples (Section~\ref{sec:measurements});
  \item validation against exact and high-precision results for ferromagnets, the three-dimensional $\pm J$ spin glass and the two-dimensional XY model (Section~\ref{sec:validation}).
\end{itemize}
Earlier versions of this article described version 0.1.8; Section~\ref{sec:corrections} lists the bugs fixed since, several of which biased results.

\subsection*{Related work}

The ALPS project~\cite{Bauer2011} provides a mature C++ framework for classical and quantum lattice models. \texttt{SpinMonteCarlo.jl}\footnote{\url{https://github.com/yomichi/SpinMonteCarlo.jl}} offers local, Swendsen--Wang, Wolff and loop updates for Ising, Potts, clock, XY and Ashkin--Teller models in Julia. Many Python repositories implement Metropolis sampling on square lattices for teaching. We are not aware of a pip-installable package that combines cluster updates, parallel tempering and replica cluster moves for spin glasses on general lattice geometries.

\section{Models}\label{sec:models}

\subsection{Ising models}

The Ising Hamiltonian is
\begin{equation}\label{eq:hamiltonian}
  H = -\sum_{\langle i,j \rangle} J_{ij}\, \sigma_i \sigma_j, \qquad \sigma_i \in \{-1, +1\},
\end{equation}
where the sum runs over nearest-neighbor pairs. A lattice is given by its shape $(L_1, \ldots, L_d)$ and a list of $n$ integer offset vectors, the forward neighbor directions, so each site has coordination number $z = 2n$. Hypercubic lattices use the $d$ unit vectors. The triangular lattice uses $\{(1,0), (0,1), (1,-1)\}$ ($n = 3$). The face-centered and body-centered cubic lattices use six and four offsets in primitive-cell coordinates ($z = 12$ and $z = 8$), so that the periodic $L_1 \times L_2 \times L_3$ index grid is a single connected lattice. These three are available as named presets through the \code{geometry} argument; any other Bravais lattice can be given through \code{neighbor\_offsets}.

The couplings $J_{ij}$ are stored on forward bonds, as an array of shape $(L_1, \ldots, L_d, n)$. They can be ferromagnetic ($J_{ij} = 1$), bimodal ($J_{ij} = \pm 1$ with equal probability), standard Gaussian, or an arbitrary user array, which can also dilute the lattice through zero entries. The lattice is periodic by default. An open or antiperiodic boundary along an axis is applied to the couplings of the bonds that cross the edge of the cell: open boundaries remove them and antiperiodic boundaries flip their sign. Unit couplings stay unit couplings, so the fast kernels described below still apply. A periodic run and an antiperiodic run on the same couplings together give the domain-wall free energy.

A simulation holds $D$ disorder samples, each with its own couplings. Each sample carries $R$ replicas of a ladder of $K$ temperatures $T_1 < T_2 < \cdots < T_K$, so the package evolves $D \times R \times K$ systems. Replicas at the same temperature share couplings, which is what overlap statistics and replica cluster moves need.

\subsection{XY models}

The XY model has unit two-component spins $\mathbf{s}_i$ and Hamiltonian
\begin{equation}
  H = -\sum_{i}\sum_{\mu=1}^{d} J_{i\mu}\, \mathbf{s}_i \cdot \mathbf{s}_{i+\hat{\mu}},
\end{equation}
on periodic hypercubic lattices in any dimension, with every extent at least three. Couplings may be ferromagnetic, bimodal, Gaussian or user-supplied; negative couplings are ordinary bonds and zero couplings are absent bonds. A Boolean occupation mask dilutes sites: a vacant site loses all its bonds and contributes nothing to magnetic or spatial observables, while all normalizations keep the full lattice size $N$. Disordered and diluted XY models of this kind are a subject of current rigorous work~\cite{Pei2026XY}. The XY model shares the disorder and replica structure, the tempering machinery and the boundary options of the Ising model, and computes in double precision.

\section{Single-system updates}\label{sec:single}

\subsection{Local updates}

\paragraph{Ising spins.} A sweep visits every site once in a fixed row-major order. The Metropolis update~\cite{Metropolis1953} flips $\sigma_i$ with probability $\min(1, e^{-\beta \Delta E})$, where $\beta = 1/T$, $\Delta E = 2\sigma_i h_i$ and $h_i = \sum_{j} J_{ij}\sigma_j$ sums over both forward and backward neighbors. The Gibbs (heat-bath) update sets $\sigma_i = +1$ with probability $(1 + e^{-2\beta h_i})^{-1}$.

Acceptance decisions use 62-bit uniform draws, so small acceptance rates are resolved rather than rounded to a single-precision grid. When all couplings lie in $\{-1, 0, 1\}$, the local field is an integer, and the kernel compares a 62-bit uniform integer with a precomputed threshold for each temperature and field value. For other couplings the Metropolis test compares $\ln u$ with $-\beta\Delta E$ for $u$ formed from the same 62-bit draw in single precision. Energies are accumulated in double precision; for unit couplings each sweep also reports its exact change of energy and magnetization. On square and simple-cubic lattices the single-spin kernels compute the neighbors of interior sites by stride arithmetic, and a sequential task advances four systems site by site in lockstep, which overlaps their dependency chains on one core without changing any trajectory.

\paragraph{XY spins.} The Metropolis update proposes a uniformly random new angle. The heat-bath update, selected by \code{sweep\_mode="gibbs"}, draws each spin exactly from the von Mises distribution $p(\theta) \propto e^{\beta |\mathbf{h}_i| \cos\theta}$ about its local field~$\mathbf{h}_i$, using the algorithm of Best and Fisher~\cite{BestFisher1979}. Unlike uniform proposals, which are rarely accepted at low temperature, it always moves the spin. Optional overrelaxation passes~\cite{Creutz1987} reflect each spin about its local field, which conserves the energy exactly.

\subsection{Fortuin--Kasteleyn cluster updates}

The Swendsen--Wang algorithm~\cite{Swendsen1987} activates each satisfied bond ($J_{ij}\sigma_i\sigma_j > 0$) with probability $1 - e^{-2\beta J_{ij}\sigma_i\sigma_j}$, builds the clusters with a union-find structure (path halving and union by rank), and flips each cluster with probability $1/2$. The Wolff algorithm~\cite{Wolff1989} grows the cluster of a random seed by depth-first search with the same bond probability and flips it.

For XY spins both algorithms run on Wolff's embedded Ising variables~\cite{Wolff1989}: a random unit vector $\mathbf{r}$ defines $a_i = \mathbf{r}\cdot\mathbf{s}_i$, a bond with $J_{ij}a_ia_j > 0$ is activated with probability $1 - e^{-2\beta J_{ij}a_ia_j}$, and flipping a cluster reflects its spins across the line orthogonal to $\mathbf{r}$, $\mathbf{s}_i \to \mathbf{s}_i - 2a_i\mathbf{r}$. Negative couplings therefore connect spins with opposite projections. This is the symmetry-based form of the construction described by Kent-Dobias and Sethna~\cite{KentDobias2018}.

A cluster update can also run in an observe-only mode, which builds the full Swendsen--Wang graph without flipping any spin and records its cluster size distribution, the moments of the cluster sizes and, on square and simple-cubic lattices, the clusters that wrap around the torus.

\section{Parallel tempering}\label{sec:pt}

Parallel tempering~\cite{Hukushima1996} exchanges configurations between neighboring temperatures. A swap between the systems at $T_k$ and $T_{k+1}$ is accepted with probability
\begin{equation}
  P_{\text{swap}} = \min\!\Big(1,\, \exp\big[(\beta_k - \beta_{k+1})(H_k - H_{k+1})\big]\Big),
\end{equation}
where $H_k$ is the total energy of the system currently at temperature $T_k$. A swap exchanges two entries of a permutation that maps temperature slots to systems, so no spins are copied.

The default schedule of the Python interface, \code{"full\_ladder"}, attempts every adjacent pair at each tempering event: first the even edges $(T_1, T_2), (T_3, T_4), \ldots$, then the odd edges. Because the order is fixed, the scheme is non-reversible~\cite{Syed2022} and carries walkers along the ladder much faster than the alternative \code{"single\_random\_edge"} schedule, which attempts one random edge per event, so that walkers move diffusively. Both schedules preserve the joint Boltzmann distribution. Each call reports the attempts and acceptances of every edge, the completed round trips of every walker between the coldest and hottest temperatures, and the flow fraction $f(T)$, the share of visits to $T$ made by walkers that last visited the cold end~\cite{Katzgraber2006}. The flow fraction falls from one to zero along the ladder, linearly when the temperatures are well spaced, and it drops steeply across a bottleneck.

\section{Replica cluster moves}\label{sec:overlap}

Replica cluster moves act on several replicas that share one set of couplings. The package implements them as instances of one construction, which also shows why some natural generalizations fail.

\subsection{Symmetries of the replica energy}

Take $M$ replicas at one temperature and write $\boldsymbol{\sigma}_i = (\sigma_i^1, \ldots, \sigma_i^M)$ for the replica spins at site~$i$; in mode names such as \code{perm}$M$, $M$ stands for this group size, as in \code{"perm4"}. A bond contributes $E_{ij}(\boldsymbol{\sigma}_i, \boldsymbol{\sigma}_j) = -J_{ij}\sum_r \sigma_i^r\sigma_j^r$ to the summed energy. Permuting the replicas, or flipping the sign of any of them, at both ends of a bond leaves its energy unchanged. Each same-temperature move picks such an involution $g$ and applies it on whole clusters of sites. The clusters are built in one of two ways.

\emph{Deterministic clusters} are the connected components of the sites that $g$ moves. Every neighbor of a component is fixed by $g$, so applying $g$ to the component conserves the summed energy exactly; the components are the same before and after the move, so the move is rejection-free and reversible.

\emph{Stochastic clusters} follow the embedding of Niedermayer~\cite{Niedermayer1988} and of Kandel and Domany~\cite{KandelDomany1991}, which treats ``apply $g$ at site $i$'' as an Ising variable. A bond is activated with probability $1 - \exp[-\beta\max(0, \Delta E_{ij})]$, where $\Delta E_{ij} = E_{ij}(\boldsymbol{\sigma}_i, g\boldsymbol{\sigma}_j) - E_{ij}(\boldsymbol{\sigma}_i, \boldsymbol{\sigma}_j)$, and the clusters are flipped as in the Swendsen--Wang or Wolff algorithm.

Every move can use Swendsen--Wang clusters (\code{"sw"}), updating every cluster (for CMR, every cluster of more than one site), or a Wolff cluster (\code{"wolff"}), updating the cluster of one random site. Same-temperature moves shuffle the replicas at each temperature into groups and update each group independently. Several moves can be alternated from one call to the next with a \code{+} in the mode name, as in \code{"gperm2+rmc"}, and the argument \code{overlap\_cluster\_max\_temperature} restricts the moves to low temperatures, where they pay off~\cite{Zhu2015}.

\subsection{Houdayer and multi-pair moves}

The Houdayer isoenergetic cluster move~\cite{Houdayer2001} takes $g$ to swap two replicas $a$ and $b$. The sites it moves are those of negative overlap, $\sigma_i^a \neq \sigma_i^b$, and the deterministic clusters are their connected components. Swapping the two replicas on a component, which is the same as flipping both, conserves the summed energy, so the move is accepted with no test and its cluster growth is independent of temperature. The move \code{pair}$M$ applies the product of $M/2$ pair swaps of a group of $M$ replicas; its clusters are the components of the sites where at least one pair disagrees.

Since the move conserves the summed energy, it must be combined with single-spin updates and parallel tempering. In three dimensions the negative-overlap sites already percolate, so flipping the dominant cluster mostly relabels the replicas; the union of several pairs percolates even more readily.

\subsection{J\"org and CMR moves}

Stochastic clusters for the same pair swap give J\"org's move~\cite{Jorg2006}: $\Delta E_{ij} = 4J_{ij}\sigma_i^a\sigma_j^a$ when both ends of the bond have negative overlap and $0$ otherwise, so satisfied bonds between negative-overlap sites are activated with probability $1 - e^{-4\beta J_{ij}\sigma_i^a\sigma_j^a}$, and the system-spanning Houdayer clusters break up. The move \code{jorg}$M$ sums $\Delta E_{ij}$ over the disagreeing pairs of a group and activates a bond only when the sum is positive.

For $g$ the flip of both replicas, $\Delta E_{ij} = 4|J_{ij}|$ on bonds that both replicas satisfy and $\Delta E_{ij} \le 0$ otherwise. These are the blue bonds of the two-replica representation of Chayes, Machta and Redner (CMR)~\cite{CMR1998,RMC1998,MachtaNewmanStein2007}, which the package implements in two phases. Write $r_{ij} = e^{-2\beta|J_{ij}|}$.
\begin{enumerate}
  \item \emph{Blue clusters.} Bonds that both replicas satisfy are activated with probability $1 - r_{ij}^2$. In Swendsen--Wang mode each blue cluster with more than one site is flipped in both replicas with probability $1/2$; in Wolff mode the seed's blue cluster is flipped.
  \item \emph{Grey clusters.} Bonds that exactly one replica satisfies, evaluated after the blue flips, are activated as red bonds with probability $1 - r_{ij}$. The grey clusters are the components of blue and red bonds together. In Swendsen--Wang mode each grey cluster with more than one site flips each replica independently with probability $1/2$; in Wolff mode the seed's grey cluster is flipped by one of the three non-trivial combinations, chosen uniformly.
\end{enumerate}
A blue flip negates both replicas, which exchanges which replica satisfies a singly satisfied bond but keeps it singly satisfied, so the red bonds can be evaluated in place on the flipped configuration. In Wolff mode, sites first reached through a red bond also sample blue bonds on their doubly satisfied edges, so that the seed's grey cluster is the full component that the Swendsen--Wang construction would give it.

Blue clusters are a proposed geometric signature of spin-glass order~\cite{MachtaNewmanStein2007}. Their percolation has been studied numerically~\cite{PeiDiVentra2021,MunsterWeigel2023,MunsterWeigel2026} and rigorously, including the number of infinite blue clusters and blue percolation in high dimensions and on hierarchical lattices~\cite{Pei2026TwoBlue,Pei2026BluePercolation,Pei2026MK}. In observe-only mode the package records blue-cluster statistics and snapshots without moving the replicas, and snapshots taken in update mode also carry the grey clusters; the repository includes a script that plots the blue-cluster size distribution of the two-dimensional $\pm J$ spin glass after Ref.~\cite{PeiDiVentra2021}.

\subsection{Replica permutations}

The move \code{perm}$M$ ($2 \le M \le 8$) uses the deterministic clusters of the sites where the $M$ replicas of a group do not all agree. Every neighbor of such a component has all replicas equal, so every replica permutation fixes it, and permuting the replicas on a component conserves the summed energy. Swendsen--Wang mode draws a uniform permutation for each cluster and Wolff mode a uniform non-identity permutation for the seed's cluster; either way the move is a rejection-free heat bath over the permutations of each cluster. The move \code{perm2} is the Houdayer move.

The gauged form \code{gperm}$M$ first picks a uniformly random reference site with replica pattern $P = (P_1, \ldots, P_M)$ and works with the gauged spins $\tau_i^r = P_r\sigma_i^r$. Sites whose pattern equals $\pm P$ are fixed, and the clusters are the components of the other sites. The probability of choosing a given gauge, the fraction of sites with pattern $\pm P$, does not change under the move, so detailed balance holds. For $M = 2$ the clusters are the sites whose overlap differs from the reference site's. That is the minority overlap sign with probability equal to the majority fraction, an exact randomized form of the minority-sign rule of Zhu, Ochoa and Katzgraber~\cite{Zhu2015}. In runs reported in the package documentation (three dimensions, $L = 8$, $T = 0.5$--$0.8$), the largest \code{gperm2} cluster covers 11--12\% of the lattice, against 48\% for the Houdayer move.

\subsection{Replica Monte Carlo across temperatures}

The move \code{rmc} pairs the systems at adjacent temperatures $T_a$ and $T_b$ within each replica, as in the replica Monte Carlo method of Swendsen and Wang~\cite{SwendsenWang1986,WangSwendsen2005}. With $\tau_i = \sigma_i^a\sigma_i^b$ held fixed, the pair is an Ising model in which flipping both systems on a connected domain of constant $\tau$ changes only the bonds on the domain wall. Such a flip is accepted with the Metropolis probability of that change, which is set by $\beta_a - \beta_b$. In Swendsen--Wang mode all $\tau = +1$ domains are decided together, then all $\tau = -1$ domains against the updated walls; neighboring domains have opposite $\tau$, so each half is a set of independent decisions. Flipping every $\tau = -1$ domain is exactly a parallel-tempering swap, so the move generalizes parallel tempering.

\subsection{A move that fails}

A tempting generalization of the Houdayer move to $M > 2$ replicas negates all replicas on clusters of \emph{balanced} sites, $\sum_r \sigma_i^r = 0$. This is not one global symmetry: a neighbor outside the cluster is not fixed by a global flip, so the boundary bonds change energy. For $J_{ij} = 1$, $\boldsymbol{\sigma}_i = (+,+,-,-)$ and $\boldsymbol{\sigma}_j = (+,+,+,-)$, flipping site~$i$ raises the summed bond energy from $-2$ to $+2$, yet the move would be accepted unconditionally. The package keeps this move, \code{houd}$M$, as an explicitly experimental option; the stationarity tests of Section~\ref{sec:validation} reject it, and in three-dimensional runs it shifts the energy and the Binder ratio far outside the errors of the valid moves.

\section{Measurements and diagnostics}\label{sec:measurements}

All observables are accumulated in double precision after a warmup fraction of the sweeps, per disorder sample and temperature. Every disorder sample's thermal averages are returned, so the user can form any disorder average and its error. A run can be split into several calls without changing its schedule, because cluster, overlap and tempering events keep their phase across calls.

\subsection{Thermodynamics and correlations}

The package records the energy and magnetization moments, the Binder cumulant~\cite{Binder1981}
\begin{equation}
  U = 1 - \frac{\langle m^4 \rangle}{3\langle m^2 \rangle^2}
\end{equation}
for Ising spins (with $2$ in place of $3$ for XY spins), and the heat capacity per spin $C = N\beta^2(\langle e^2\rangle - \langle e\rangle^2)$, where $e = H/N$. The heat capacity takes the thermal variance within each disorder sample before averaging over samples, so it excludes the disorder variance of the thermal mean energy. For backward compatibility, the Ising result key \code{energies} holds the interaction sum $-H/N$, while the physics collector and the XY model report $H/N$. An optional physics collector adds structure factors $S(\mathbf{k})$ at the smallest nonzero wavevector along each axis, the second-moment correlation length~\cite{CooperFreedmanPreston1982}
\begin{equation}
  \xi_\mu = \frac{\sqrt{S(0)/S(k_\mu) - 1}}{2\sin(k_\mu/2)}, \qquad k_\mu = \frac{2\pi}{L_\mu},
\end{equation}
spin correlations at requested displacements, block sums for error analysis, and delete-one-sample jackknife errors for every disorder-averaged quantity, including ratios of averages. For XY spins it also records the helicity modulus
\begin{equation}
  Y_\mu = \frac{\langle D_\mu\rangle - \beta\langle I_\mu^2\rangle}{N}, \qquad D_\mu = \sum_i J_{i\mu}\,\mathbf{s}_i\cdot\mathbf{s}_{i+\hat\mu}, \qquad I_\mu = \sum_i J_{i\mu}\,\mathbf{s}_i\times\mathbf{s}_{i+\hat\mu},
\end{equation}
and, in two dimensions, the vortex density.

\subsection{Overlaps}

With $R \ge 2$ replicas the package measures the site overlap and the link overlap of replica pairs,
\begin{equation}
  q = \frac{1}{N}\sum_{i} \sigma_i^{a} \sigma_i^{b}, \qquad
  q_l = \frac{1}{N_b}\sum_{\langle i,j \rangle} \sigma_i^{a} \sigma_j^{a}\sigma_i^{b} \sigma_j^{b},
\end{equation}
where the sum runs over all $N_b = nN$ forward neighbor pairs of the periodic lattice, including pairs whose coupling is zero or removed by an open boundary. The moments $\langle q^2\rangle$ and $\langle q^4\rangle$ use the fixed pairs $(1,2), (3,4), \ldots$ and give the spin-glass Binder ratio $g = 1 - [\langle q^4 \rangle]/(3[\langle q^2 \rangle]^2)$, where $[\cdot]$ is the disorder average (equivalently $[\langle q^4\rangle]/[\langle q^2\rangle]^2 = 3(1-g)$); its crossings for different sizes locate $T_c$~\cite{BaityJesi2013}. The full overlap distribution $P(q)$ is accumulated exactly, one bin per possible value of $Nq$, together with the link overlap conditioned on $q$. The physics collector adds the overlap structure factor at the smallest wavevectors and from it the spin-glass correlation length $\xi_{\mathrm{SG}}$ and $\xi_{\mathrm{SG}}/L$~\cite{Ballesteros2000}, defined as above with $\chi_{\mathrm{SG}}(\mathbf{k})$ in place of $S(\mathbf{k})$. It also records the replica-symmetric
\begin{equation}
  Q_2 = \binom{R}{2}^{-1}\sum_{a<b} q_{ab}^2,
\end{equation}
which, unlike $q^2$ of a fixed pair, does not change when replicas are relabeled.

\subsection{Autocorrelation times}

The integrated autocorrelation time of $m^2$, of the energy and of $Q_2$ is estimated with Sokal's automatic window~\cite{Sokal1997},
\begin{equation}
  \tau_{\mathrm{int}} = \frac{1}{2} + \sum_{w=1}^{W} \Gamma(w),
\end{equation}
where $\Gamma$ is the normalized autocorrelation function and $W$ is the smallest window with $W \ge 5\tau_{\mathrm{int}}(W)$. The default backend accumulates the autocorrelation exactly with memory proportional to the maximum lag; an FFT backend keeps the full series instead. Each estimate carries a flag saying whether the window closed within the available lags; when it did not, $\tau_{\mathrm{int}}$ is a truncated lower bound. A series without variance reports no value. The overlap time uses $Q_2$, so moves that only relabel replicas do not count as decorrelation. Under full-ladder tempering the autocorrelation of $Q_2$ has a small first lag and a long tail, and the automatic window can underestimate $\tau_{\mathrm{int}}$; Table~\ref{tab:moves} therefore uses batch means.

\subsection{Equilibration}

For Gaussian couplings of variance $J^2$ on a periodic lattice, integrating by parts over the disorder gives the equilibrium identity $-[\langle H\rangle]/N = J^2\beta\,(N_b/N)\,(1 - [\langle q_l\rangle])$~\cite{KPY2001}. The package evaluates
\begin{equation}
  \Delta(t) = -\frac{H(t)}{N} - J^2\beta\,\frac{N_b}{N}\,\bigl(1 - q_l(t)\bigr),
\end{equation}
with $N_b/N = n$, averaged over replicas, disorder samples and the logarithmic windows $[t/2, t)$ for $t = 128, 256, 512, \ldots$ sweeps. From a random start both terms rise toward equilibrium, so $\Delta$ approaches zero from below; a simulation is considered equilibrated once $\Delta$ agrees with zero within errors over the last few windows~\cite{Zhu2015}. The package returns the disorder average of $\Delta$; its error requires repeated or split runs. The identity does not hold for bimodal couplings. For those, the package records $Q_2$ in the same windows, and a plateau of $Q_2$ across windows serves as the criterion instead.

\subsection{Move diagnostics}

Fortuin--Kasteleyn updates count the spins they visit; for a ferromagnet the mean Wolff cluster size estimates $N\langle m^2\rangle$, which converts autocorrelation times measured in cluster steps into sweeps. Replica cluster moves count, per mode and temperature, their group updates, the sites they act on and the sites eligible for the move; a move that acts on more than half of the eligible sites mostly relabels replicas. Replica Monte Carlo counts proposed and accepted domains and flipped sites per temperature edge. Overlap cluster size distributions and the relative sizes of the four largest clusters are also available.

\section{Implementation}\label{sec:implementation}

\subsection{Architecture}

The repository contains a pure Rust crate, \code{spin-sim}, a thin PyO3 extension module, and a Python layer. The crate holds the lattice geometry, the update kernels, the cluster and replica moves, the statistics, and a simulation driver that schedules sweeps, cluster updates, replica moves, measurements and tempering. A generic realization type, parameterized by the spin type, holds the spins, couplings, random number generators and tempering state of one disorder sample; Ising realizations use \code{i8} spins and \code{f32} couplings, and XY realizations use pairs of \code{f64}. The Python classes \code{Ising} and \code{XY} prepare couplings and temperatures, call the extension, and post-process the results. A command-line interface, \code{peapods}, runs single simulations, parameter sweeps and benchmarks.

\subsection{Data layout}

Spins of all replicas and temperatures of a realization are stored in one flat array, and couplings only on forward bonds; the coupling of a backward bond is the forward coupling of the neighboring site. The lattice precomputes, for each of the $N$ sites and $n$ directions, the forward and backward neighbor indices, a table of $2nN$ \code{u32} entries: 96\,KB for a $64\times 64$ triangular lattice, 6\,MB for a $64^3$ cubic lattice and 12\,MB for a $64^3$ FCC lattice. Single-spin sweeps on square and simple-cubic lattices use stride arithmetic for interior sites; all other kernels read the table. Spin, coupling, generator and measurement buffers are allocated once; cluster updates allocate scratch storage per update, and replica moves draw theirs from a per-thread pool.

\subsection{Parallelism}

Systems are independent given the temperature assignment, so sweeps, energy updates and cluster moves run in parallel over systems with Rayon's work-stealing scheduler~\cite{Rayon}; each task receives a disjoint slice of the spin array and its own random number generator. Disorder samples run in parallel as well, and once they fill the thread pool each sample processes its replicas on its own thread instead of nesting parallelism. Tempering is sequential within a sample and cheap. The extension releases Python's global interpreter lock while sampling and can be interrupted from the keyboard.

\subsection{Random numbers and reproducibility}

Every system and every replica-pair slot has its own Xoshiro256** generator~\cite{BlackmanVigna2021}, seeded from one root seed through domain-separated SplitMix64 hashes~\cite{Steele2014}. The \code{seed} argument makes the couplings and the dynamics reproducible, and \code{reset()} restarts the dynamics while keeping the couplings.

\subsection{Testing}

The Rust test suite enumerates every state of a $3\times 2$ torus and compares the histogram after one update with the Boltzmann law, by a $\chi^2$ test, for the Metropolis and Gibbs kernels (lookup and generic), the Swendsen--Wang and Wolff updates and the tempering exchange; in a one-off check during development, a sweep with a $10\%$ error in $\beta$ gave $\chi^2 = 662$ against a cutoff of about $119$. The replica moves have their own one-step tests (Section~\ref{sec:validation}). Python tests cover the interfaces, the statistics, and thermodynamic consistency, for instance that the heat capacity of a spin glass integrates over temperature to the change of its energy. Continuous integration runs these tests, the lints and the bounded validations of Section~\ref{sec:validation} on every change.

\section{Validation}\label{sec:validation}

\subsection{Exact stationarity of the replica moves}

For each replica move, the test draws the replicas exactly from the Boltzmann law on a small torus, applies one move, and compares the histograms of states, the energies and the pairwise overlap histograms with the product law. The test runs at $T = 3$ on $2\times 2$ and $3\times 2$ tori, with $40{,}000$ trials and a $5\sigma$ tolerance, in both cluster modes and, for most moves, with and without statistics collection. It covers the Houdayer, J\"org and CMR moves; \code{pair}$M$ and \code{jorg}$M$ for $M = 2, 4, 6$; \code{perm}$M$ and \code{gperm}$M$ for $M = 2$ to $5$; and replica Monte Carlo on ladders of two and three temperatures. It detects wrong bond probabilities, a naive single-pair rule for \code{jorg}$M$, wrong replica Monte Carlo couplings, deciding all replica Monte Carlo domains at once, a missing gauge when \code{gperm}$M$ writes spins back, the incomplete grey cluster of CMR's Wolff mode in version 0.2.0 (Section~\ref{sec:corrections}), and the balanced-site \code{houd4} move.

\subsection{Ising ferromagnets}

\begin{figure}[ht]
  \centering
  \includegraphics[width=0.65\textwidth]{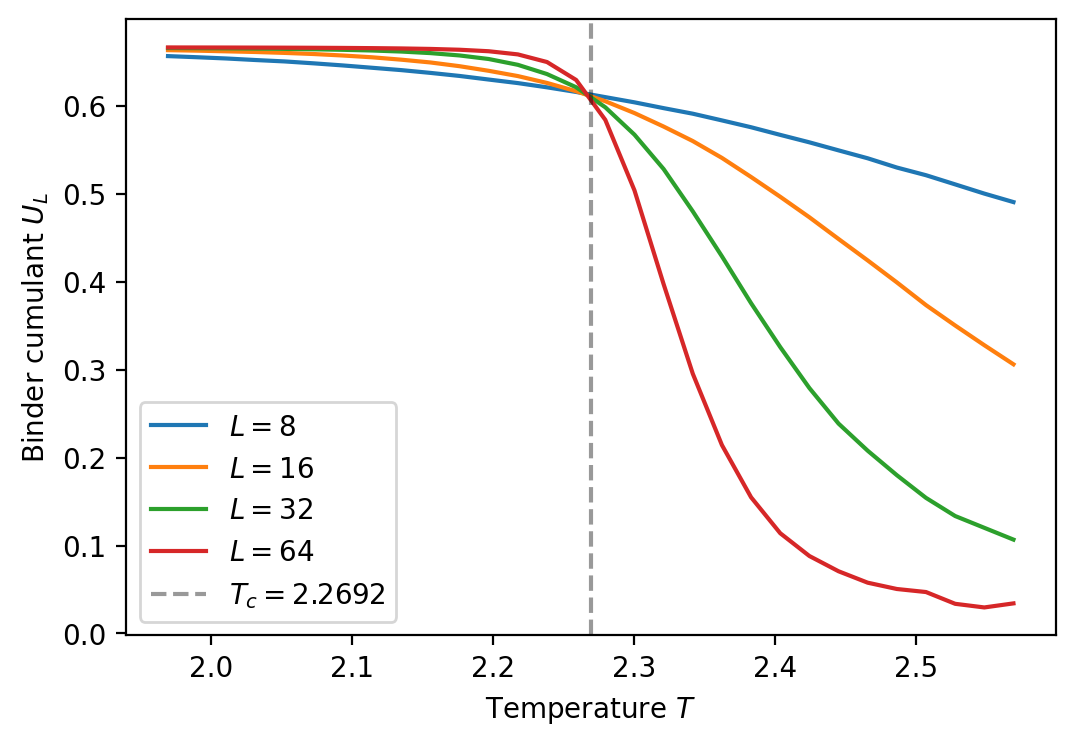}
  \caption{Binder cumulant $U_L$ against temperature for the square-lattice Ising ferromagnet, $L = 8, 16, 32, 64$, from 500{,}000 sweeps of Metropolis plus Wolff updates with parallel tempering. The dashed line marks the exact $T_c \approx 2.269$. These data were produced with version 0.1.8 for an earlier version of this article; the current code passes the same check at smaller sizes in continuous integration.}
  \label{fig:binder}
\end{figure}

Figure~\ref{fig:binder} shows the Binder cumulant of the square-lattice ferromagnet for $L = 8$ to $64$. The curves cross at $T \approx 2.27$ with $U \approx 0.61$, consistent with the exact $T_c = 2/\ln(1+\sqrt{2}) \approx 2.269$~\cite{Onsager1944} and the universal crossing value of the two-dimensional Ising class. The same analysis with version 0.1.8 on the triangular lattice, a test of the general neighbor offsets, gave a crossing at $T \approx 3.64$, against the exact $T_c = 4/\ln 3 \approx 3.641$~\cite{Wannier1950,Houtappel1950}. Continuous integration repeats these checks with the current code at $L = 8$, $16$ and $32$, and extends them to the simple-cubic, BCC and FCC lattices at $L = 6$, $8$ and $10$. It requires the Binder curves of all sizes to agree within $0.05$ at the literature critical temperatures, $\beta_c = 0.2216546$, $0.1573725$ and $0.1020707$ respectively; the first two agree with the series estimates $0.221654(1)$ and $0.1573725(6)$ of Butera and Comi~\cite{ButeraComi2000}, and the FCC value is from Ref.~\cite{Lundow2009}.

\subsection{Three-dimensional \texorpdfstring{$\pm J$}{+-J} spin glass}

As a test of the spin-glass machinery we reproduce the finite-size crossings that locate the transition of the three-dimensional Edwards--Anderson model with $\pm J$ couplings, for which the Janus collaboration reports $T_c = 1.1019(29)$, $\xi/L = 0.6516(32)$ and $U_4 = 1.4899(28)$ at $T_c$~\cite{BaityJesi2013}. We simulated periodic $L^3$ lattices with $L = 4, 5, 6, 8, 10, 12$, $3840$ disorder samples per size and two replicas per sample, at 31 temperatures from $1.0$ to $1.6$. Each sweep was one Metropolis sweep of all systems followed by one full-ladder tempering pass. Runs lasted $2^{14}$ sweeps for $L = 4, 5$, $2^{15}$ for $L = 6$, $2^{16}$ for $L = 8$, $2^{17}$ for $L = 10$ and $2^{18}$ for $L = 12$, at least eight times the equilibration time measured in a pilot run, and results use the second half of each run. In every size the last two logarithmic windows agree at $T = 1.0$, the lowest temperature, within $1.5$ paired standard errors in $\xi_{\mathrm{SG}}/L$ and $1.7$ in $U_4$. The run took $5.3$ hours on 96 cores. It used the development code of early October 2026, before two changes to tempering: a fixed order of the even and odd passes, which speeds up the walk along the ladder, and 62-bit uniforms in the swap decision. Neither changes the sampled distribution at the precision reported; the smallest swap acceptance in the run was $0.79$.

\begin{figure}[ht]
  \centering
  \includegraphics[width=\textwidth]{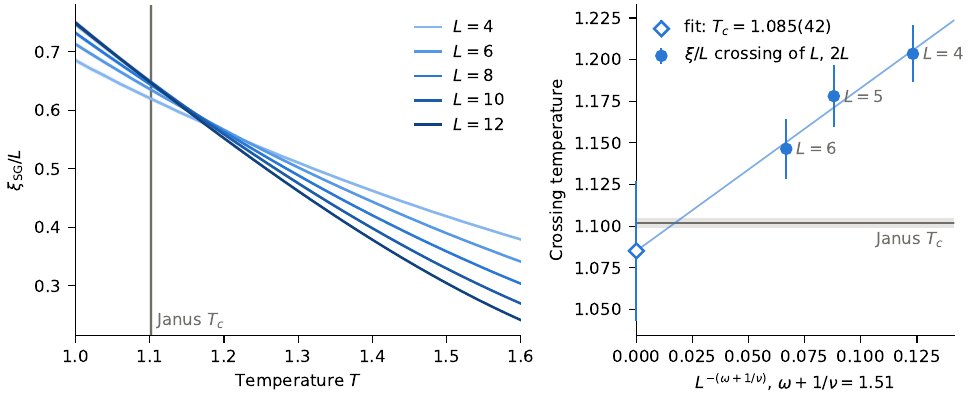}
  \caption{Three-dimensional $\pm J$ spin glass. Left: spin-glass correlation length $\xi_{\mathrm{SG}}/L$ against temperature for $L = 4$ to $12$ ($L = 5$ omitted for clarity; shading shows one standard error); the vertical line is the Janus $T_c$. Right: crossing temperatures of $L$ and $2L$ against $L^{-(\omega+1/\nu)}$ with $\omega + 1/\nu = 1.51$, the weighted fit, and its intercept $T_c = 1.085(42)$; the horizontal band is the Janus value $1.1019(29)$~\cite{BaityJesi2013}.}
  \label{fig:sg}
\end{figure}

The physics collector gives per sample $\langle q^2\rangle$, $\langle q^4\rangle$ and $\chi_{\mathrm{SG}}(k_{\min})$, from which $\xi_{\mathrm{SG}}/L$ and $U_4 = [\langle q^4\rangle]/[\langle q^2\rangle]^2 = 3(1-g)$ follow. Errors are delete-one-block jackknife estimates over 64 blocks of disorder samples. A crossing is the highest temperature at which the curves of $L$ and $2L$ intersect, by linear interpolation on the temperature grid. Table~\ref{tab:sg} lists the crossings, and Figure~\ref{fig:sg} shows $\xi/L$ and the extrapolation. Fitting $T^*(L, 2L) = T_c + a\,L^{-(\omega + 1/\nu)}$ with the Janus value $\omega + 1/\nu = 1.51$ gives $T_c = 1.085(42)$ with $\chi^2 = 0.21$ for one degree of freedom, consistent with the Janus value. As expected on lattices this small, the crossings drift down and $\xi_{\mathrm{SG}}/L$ at the crossing drifts up toward $0.652$. The extrapolation leans on the fixed correction exponent, so it is a consistency check of the sampler rather than an independent determination of $T_c$.

\begin{table}[ht]
  \centering
  \begin{tabular}{cccc}
    \toprule
    $(L, 2L)$ & $T^*$ from $\xi_{\mathrm{SG}}/L$ & $\xi_{\mathrm{SG}}/L$ at $T^*$ & $T^*$ from $U_4$ \\
    \midrule
    $(4, 8)$  & $1.204(17)$ & $0.561$ & $1.130(37)$ \\
    $(5, 10)$ & $1.178(19)$ & $0.578$ & $1.138(28)$ \\
    $(6, 12)$ & $1.146(18)$ & $0.603$ & $1.107(30)$ \\
    \bottomrule
  \end{tabular}
  \caption{Crossing temperatures $T^*$ of the curves for $L$ and $2L$ in the three-dimensional $\pm J$ spin glass.}
  \label{tab:sg}
\end{table}

Continuous integration runs a smaller check of the same model with each of the Houdayer, J\"org, CMR and replica Monte Carlo moves: at $L = 8$ and $10$ the two curves of $g$ must agree within $0.3$ at the Janus $T_c$, and both must lie between $0.3$ and $0.7$ there, a band that a grossly wrong sampler, such as \code{houd4} with $g \approx 0$, would leave.

\subsection{Two-dimensional XY model}

At $\beta = 1.1199$, close to the Berezinskii--Kosterlitz--Thouless transition, Hasenbusch~\cite{Hasenbusch2005} gives high-precision values of the helicity modulus $\beta Y$, the second-moment correlation length $\xi/L$ and the structure factor $S(0)$. A bounded validation, run in continuous integration, starts four independent chains per size with $4096$ warmup and $16{,}384$ measured sweeps of Metropolis and embedded Swendsen--Wang updates. It requires agreement within four combined standard errors and statistical errors below 1\% of each value. Table~\ref{tab:xy} shows a run of 3 October 2026: all six comparisons pass, with the largest deviation $1.86$ combined standard errors. All six estimates lie slightly below the reference values; they are correlated, because they come from the same chains.

\begin{table}[ht]
  \centering
  \begin{tabular}{lccc}
    \toprule
    & $\beta Y_0$ & $\xi_0/L$ & $S(0)$ \\
    \midrule
    $L = 16$, \pkg{}       & $0.72282(144)$ & $0.79928(401)$ & $132.743(196)$ \\
    $L = 16$, Ref.~\cite{Hasenbusch2005} & $0.72536$ & $0.79953$ & $133.011$ \\
    $L = 32$, \pkg{}       & $0.70803(121)$ & $0.78382(459)$ & $450.857(674)$ \\
    $L = 32$, Ref.~\cite{Hasenbusch2005} & $0.70883$ & $0.79231$ & $452.114$ \\
    \bottomrule
  \end{tabular}
  \caption{Two-dimensional XY model at $\beta = 1.1199$: helicity modulus and correlation length along the first axis, and structure factor $S(0)$, against the reference values of Hasenbusch~\cite{Hasenbusch2005}.}
  \label{tab:xy}
\end{table}

\section{Performance}\label{sec:performance}

Table~\ref{tab:bench} gives the wall time per sweep of a $64\times 64$ ferromagnet with 16 temperatures, geometrically spaced from $0.1$ to $10$, and 128 disorder samples, $8.4\times 10^6$ spins in all, over runs of 50 sweeps\footnote{Script \code{benchmarks/sweep\_modes.py} at commit \code{c6404e2} of the repository; the July build is commit \code{d945d92}.}, measured on a 24-thread desktop CPU (Intel Core Ultra 9 285K, Linux). Each sweep includes the measurements. The last column is the same benchmark for a development build of July 2026, before the kernel changes of Section~\ref{sec:single}; local sweeps are 2.4--3.5 times faster. The tempering row used the single-random-edge schedule, then the default. It does strictly more work than the Metropolis row yet measured faster, so differences of about 10\% between rows are within run-to-run variation.

\begin{table}[ht]
  \centering
  \begin{tabular}{lcc}
    \toprule
    Updates per sweep & September 2026 (ms) & July 2026 build (ms) \\
    \midrule
    Metropolis & $1.47$ & $3.49$ \\
    Gibbs & $1.37$ & $4.83$ \\
    Metropolis + Swendsen--Wang & $7.63$ & $14.82$ \\
    Metropolis + Wolff & $3.99$ & $7.10$ \\
    Metropolis + parallel tempering & $1.35$ & $3.19$ \\
    \bottomrule
  \end{tabular}
  \caption{Wall time per sweep, median of five interleaved runs, for a $64 \times 64$ ferromagnet with 16 temperatures and 128 disorder samples. Measured on 29 September 2026; later changes were not re-benchmarked on this machine.}
  \label{tab:bench}
\end{table}

Whether a replica cluster move pays depends on the system. Table~\ref{tab:moves} compares, for $\pm J$ spin glasses with four replicas, Metropolis sweeps and full-ladder tempering every sweep, and 32 disorder samples, the CPU time per sweep multiplied by the integrated autocorrelation time of $Q_2$, relative to tempering alone; values below one mean the move pays. At these sizes tempering alone is hard to beat. From a random start, however, the Houdayer and gauged moves applied every fourth sweep reach the energy and $Q_2$ that tempering alone reaches after $65{,}536$ sweeps about 25 and at least 6 times faster in two dimensions at $L = 32$ and $T = 0.20$, and about $1.5$ times faster in three dimensions at $L = 16$ and $T = 0.40$; there is no gain in three dimensions at $L = 12$ and $T = 0.40$, or at $L = 16$ and $T = 0.60$. These ratios are single measurements without error estimates. Moves on more than two replicas are valid; at these sizes they do no better than the best two-replica move, and their clusters are larger: in the runs reported in the package documentation, the largest \code{perm4} cluster covers 88\% of the lattice at $L = 8$ in three dimensions at every temperature.

\begin{table}[ht]
  \centering
  \begin{tabular}{lccccc}
    \toprule
    System, temperature & $\tau$, tempering only & \code{houd2} & \code{gperm2} & \code{rmc}, every 4th sweep & \code{perm4} \\
    \midrule
    2D, $L = 16$, $T = 0.30$ & 14 & 1.5 & 1.5 & 1.4 & 1.6 \\
    3D, $L = 8$, $T = 0.50$  & 25 & 2.3 & 1.6 & 2.1 & 1.9 \\
    3D, $L = 8$, $T = 0.80$  & 15 & 2.4 & 2.4 & 2.1 & 2.5 \\
    \bottomrule
  \end{tabular}
  \caption{Cost of replica cluster moves for $\pm J$ spin glasses with 16 temperatures: CPU time per sweep times $\tau_{\mathrm{int}}(Q_2)$, relative to parallel tempering alone, whose $\tau_{\mathrm{int}}$ in sweeps is given in the second column. \code{houd2}, \code{gperm2} and \code{perm4} act every sweep in Wolff mode. Autocorrelation times are from batch means of 1000 sweeps.}
  \label{tab:moves}
\end{table}

\section{Usage}\label{sec:usage}

Listing~\ref{lst:usage} builds a ferromagnet, a triangular lattice with an open boundary, a three-dimensional spin glass with a gauged replica move below $T_c$, a Gaussian spin glass with the equilibration diagnostic, and a diluted XY model.

\begin{lstlisting}[float=tbp, caption={Example use of the Python interface of version 0.3.0.}, label=lst:usage]
import numpy as np
from peapods import Ising, XY

# 2D ferromagnet: Metropolis + Swendsen-Wang + full-ladder tempering
model = Ising((32, 32), temperatures=np.linspace(2.0, 2.6, 16), seed=1)
model.sample(10_000, cluster_update_interval=1, cluster_mode="sw",
             pt_interval=1)
print(model.binder_cumulant)

# Triangular lattice, open along the second axis
tri = Ising((32, 32), geometry="triangular",
            temperatures=np.linspace(3.2, 4.0, 16),
            boundaries=("periodic", "open"))
tri.sample(10_000, cluster_update_interval=1, pt_interval=1)

# 3D +-J spin glass: 64 disorder samples with 4 replicas each
sg = Ising((8, 8, 8), couplings="bimodal", n_disorder=64,
           n_replicas=4, temperatures=np.geomspace(0.5, 1.6, 24),
           seed=7)
result = sg.sample(2**16, pt_interval=1,
                   overlap_cluster_update_interval=4,
                   overlap_cluster_build_mode="gperm2",
                   overlap_cluster_max_temperature=1.1,
                   collect_physics=True, autocorrelation_max_lag=1024)
physics = result["physics"]
print(sg.sg_binder, sg.overlap2_tau)
print(physics["sg_correlation_length_ratio"],
      physics["stderr"]["sg_correlation_length_ratio"])
print(result["per_disorder"]["parallel_tempering"]["round_trips"])

# Gaussian couplings: equilibration diagnostic Delta(t)
gauss = Ising((8, 8, 8), couplings="gaussian", n_disorder=64,
              n_replicas=2, temperatures=np.geomspace(0.5, 1.6, 24))
gauss.sample(2**16, pt_interval=1, equilibration_diagnostic=True)
sweeps, delta = gauss.equilibration_delta()

# 2D XY with 10% site dilution and an exact heat bath
occupation = np.random.default_rng(0).random((32, 32)) > 0.1
xy = XY((32, 32), temperatures=[0.7, 0.9, 1.1],
        occupation=occupation, seed=3)
r = xy.sample(4096, sweep_mode="gibbs", pt_interval=1)
print(r["helicity_modulus"])
\end{lstlisting}

The documentation at \url{https://peabrane.github.io/peapods/} describes every argument and result key, with guides to the replica moves and the XY model.

\section{Availability and earlier versions}\label{sec:corrections}

\pkg{} is released under the MIT license at \url{https://github.com/PeaBrane/peapods}. Version 0.3.0 is installed with \code{pip install peapods}; wheels are built for CPython 3.11 to 3.13 on Linux (x86\_64 and aarch64), macOS (Intel and Apple silicon) and Windows (x86\_64), and the Rust core is published separately as the crate \code{spin-sim}. The validation scripts and the data behind Figure~\ref{fig:sg} are in the repository's \code{validation} directory.

Earlier versions of this article described version 0.1.8. The following bugs, present in version 0.2.0, are fixed in 0.3.0:
\begin{itemize}
  \item The \code{bcc} and \code{fcc} presets used conventional-cell offsets on the cubic index grid, which split lattices of even size into four (BCC) or two (FCC) disconnected copies.
  \item In Wolff mode without cluster statistics, the defaults for replica moves, the CMR move grew the grey cluster from the flipped blue cluster through red bonds only and never absorbed blue bonds leaving sites reached through a red bond. The move violated detailed balance and biased the energy.
  \item The heat capacity included the disorder variance of the thermal mean energy.
  \item The autocorrelation function was not exactly centred, which biased $\tau_{\mathrm{int}}$ when the mean of a series is large compared with its spread, as for $m^2$ below $T_c$.
  \item Acceptance tests drew single-precision uniforms, so every proposal was accepted with probability at least $2^{-24}$.
  \item A lattice extent of 1 made a site its own neighbor and sampled a wrong distribution; such lattices are now rejected.
  \item Overlap cluster-size histograms counted sites outside the move as single-site clusters.
\end{itemize}
Results obtained before version 0.3.0 with BCC or FCC lattices, with the CMR move in Wolff mode, or for heat capacities of disordered models should be recomputed, and autocorrelation times of observables with a large mean re-estimated. Other changes alter definitions or defaults: the equilibration diagnostic averages over the windows $[t/2, t)$ instead of from the first sweep; parallel tempering defaults to the full-ladder schedule; the Swendsen--Wang variants of the Houdayer and J\"org moves also update single-site clusters; quantities that were not measured are reported as undefined rather than zero; and interval events keep their phase across calls. Finally, the equilibration diagnostic in earlier versions of this article had the coordination number $z$ where the bond density $N_b/N = z/2$ belongs, and did not state that the identity behind it holds only for Gaussian couplings; the code always used $N_b/N$.

\section{Conclusion}

\pkg{} combines a Python interface with a Rust sampling loop for Ising and XY models with arbitrary couplings. It provides the standard local, cluster and tempering updates; a family of replica cluster moves for spin glasses, the same-temperature ones derived from one symmetry construction, all checked by stationarity tests against exact Boltzmann samples; and the measurements needed to judge equilibration and to compare algorithms. Its results agree with exact and literature references for Ising ferromagnets, the three-dimensional $\pm J$ spin glass and the two-dimensional XY model.

Planned extensions include Potts, clock and Heisenberg spins with their disordered variants, cluster algorithms for frustrated systems such as that of Kandel, Ben-Av and Domany~\cite{KBD1990}, external fields, and, for XY spins, custom lattice geometries, spin-glass observables and chirality.

\bibliographystyle{unsrt}
\bibliography{refs}

\end{document}